\documentclass{elsart}

\usepackage{epsfig}

\def\bea {\begin{eqnarray}}
\def\eea {\end{eqnarray}}

\def\be {\begin{equation}}
\def\ee {\end{equation}}
\begin{document} % do not change

\begin{frontmatter} % do not change

\title{Probing collectivity in ultra-relativistic heavy ion collision 
by leptons and photons}

\author{Jan-e Alam}
\medskip
\address{Theoretical Physics Division, Variable Energy Cyclotron Centre, 
1/AF, Bidhan Nagar, Kolkata - 700064, India}
%\date{\today}

%=====================================================================
\begin{abstract}
It has been shown that the evolution of collectivity in 
ultra-relativistic heavy ion collision is manifested in
the variation of various HBT radii with invariant mass ($M$)
extracted from the correlation functions of two lepton pairs.
The value of the radial velocity ($v_r$) can be estimated 
from the ratio of the $p_T$ distributions of single photons to 
lepton pairs for various $M$ windows. It has been argued that 
the variation of radial flow with appropriate kinematic 
variables can be used as an indicator of a phase transition from initially 
produced partons to hadrons. We also consider 
the elliptic flow ($v_2^{HF}$) of the matter as probed by the
single electron spectra originating from the semileptonic decays of heavy 
mesons. The measured values of $v_2^{HF}$ and the nuclear suppression factor
($R_{\mathrm AA}$) at RHIC energy have been reproduced simultaneously 
by including both the collisional and radiative processes within 
the scope of perturbative quantum chromodynamics. The $R_{\mathrm AA}$
and $v_2^{HF}$ have been predicted for LHC energy.
\end{abstract}
%=======================================================================
%\maketitle

\end{frontmatter} % do not change

\section{Introduction}
The hot and dense matter expected to be formed in the partonic phase after 
ultra-relativistic heavy ion collisions (URHIC) dynamically evolve in space and time 
due to high internal pressure. The system cools due to expansion
and reverts back to hadronic matter from the partonic phase. 
%Just after the formation,
%the entire energy of the system is thermal in nature
%and with progress of time some
%part of the thermal energy gets converted to the collective (flow) energy.
%In other words, during the expansion stage
%the total energy of the system is shared by the thermal
%as well as the collective degrees of freedom. 
%The evolution of
%the collectivity within the system is sensitive to the Equation of State
%(EoS) and hence on the parton-hadron transition.
It is well known that the average magnitude of
radial flow at the freeze-out surface can be extracted from
the transverse momentum ($p_T$) spectra of the hadrons.
However, hadrons being strongly interacting objects
can bring the information of the state of the system
when it is too dilute to support collectivity.
% {\it i.e.}
%the parameters of collectivity extracted from the hadronic
%spectra are limited to the evolution stage where the
%collectivity ceases to exist. These collective parameters have
%hardly any information about the interior of the matter.
The electromagnetic (EM) probes, {\it i.e.}
photons and dileptons on the other hand are produced and emitted~\cite{lm}
(see~\cite{ja} for review) from each space time points.
Therefore, estimating radial flow from the EM probes will shed light on
the time evolution of the collectivity in the system. The generation of
collectivity in the system depends on EoS - hence this can be used to
differentiate partonic and hadronic phases as the EoS for these two
phases are different.  
In case of EM probes- dilepton has the advantage over the real photons.
Because the low $p_T$ photons from the
hadronic phase receive large transverse kick
due to  radial flow and consequently appear in the high $p_T$ domain to 
mingle with  those from the QGP phase,
making the detection of photons from QGP difficult.
However, for dileptons there are two kinematic variables available - the
$p_T$ and the invariant mass ($M$).
While the $p_T$ spectra of dilepton is affected by the flow,
the $p_T$ integrated $M$ spectra remains unaltered.
This suggests that a careful selection of $p_T$ and
$M$ windows will be very useful to characterize the
QGP and hadronic phases.  In the present work
we will demonstrate how the development of radial flow can be
estimated through the Hanbury-Brown Twiss (HBT) interferometry
with virtual photons (lepton pairs) for different $M$ windows. 
The radial flow velocity ($v_r$) can be estimated by considering the
ratio of the $p_T$ distribution of single photon to lepton pairs
for various $M$ windows. We will briefly discuss the results here and
refer to~\cite{pm} for details.

Single electrons originating from the semi-leptonic 
decays of heavy mesons carry the information on the interaction of the
heavy quarks (a constituent of the heavy mesons) with the  
thermal medium of light quarks and gluons produced in heavy ion collisions.
The $R_{\mathrm AA}$ and $v_2^{HF}$ can be used to quantify 
the interaction of the heavy quarks with the QGP.
Several ingredients like inclusions of non-perturbative
contributions from the quasi-hadronic bound state~\cite{hvh},
3-body scattering effects~\cite{ko},
the dissociation of heavy mesons due to its
interaction with the partons in the
thermal medium~\cite{adil}  and employment of running coupling
constants and realistic Debye mass~\cite{gossiaux},
the inclusion of both elastic and inelastic collisions  
along with the path length fluctuation
have been proposed~\cite{wicks} to improve the description of the
experimental data.  Within the framework of Fokker Planck equation (FPE) 
we will evaluate $v_2^{HF}$ and $R_{\mathrm AA}$  for these electrons.
In the present paper we discuss the elliptic flow of the matter
probed by the single electron from the heavy mesons decays. 
For the elliptic flow of the matter probed by single photon and lepton pair
we refer to~\cite{rc1,rc2} for details.
 
In the next section we will briefly describe the HBT interferometry
with virtual photons. The azimuthal anisotropy of the
system probed by single electrons originating from the heavy flavour decays
will be discussed in section 3. Section 4 is devoted to summary and 
discussions.

\section{HBT interferometry with dileptons}
The interferometry of the dilepton pairs actually reflect correlation
between two virtual photons, the analysis then can proceed by
computing the Bose-Einstein correlation (BEC) function for two 
virtual photons which can be defined as,
$C_{2}(\vec{p_{1}}, \vec{p_{2}}) = P_{2}(\vec{p_{1}}, \vec{p_{2}})/
\left[P_{1}(\vec{p_{1}}) P_{1}(\vec{p_{2}})\right]$,
where $p_i$ is momentum of the individual lepton pair,
$P_{1}(\vec{p_{i}})$
and $P_{2}(\vec{p_{1}}, \vec{p_{2}})$ represent
the one- and two- particle inclusive
lepton pair spectra respectively,which can be evaluated form the
source function for various invariant mass windows of the pair~\cite{mam}.
For the productions of lepton pairs from QGP the annihilation of
thermal quarks and  from the hadronic phase the decays of thermal light
vector mesons ($\rho$, $\omega$ and $\phi$) have been considered. 

For the space time evolution of the system relativistic hydrodynamical
model with cylindrical symmetry~\cite{hvg} and  boost invariance along
the longitudinal direction~\cite{jdb} has been used. 
For a system undergoing isentropic expansion,
the initial 
temperature ($T_{i}$) and proper thermalization time 
($\tau_{i}$) of the system 
may be constrained by the measured hadronic multiplicity,
$dN/dy\sim T_i^3\tau_i$. 
For Relativistic Heavy Ion Collider (RHIC)
we have taken $T_i=290$ MeV and $\tau_i=0.6$ fm/c.
The EoS which controls the rate of expansion/cooling
has been taken from the lattice QCD calculations ~\cite{MILC}. 
The chemical ($T_{ch}$=170 MeV) and kinetic ($T_{f}$=120 MeV)
freeze-out temperatures 
are fixed by the particle ratios and the slope 
of the $p_T$ spectra of hadrons~\cite{hirano}. 
With all these ingredients the correlation function $C_2$ %(Fig~\ref{fig1})
has been  evaluated for different (average) invariant mass windows,
$\langle M \rangle=(M_1+M_2)/2$ 
as a function of $q_{side}$ and $q_{out}$~\cite{mam} which are related 
to the transverse momentum of individual pair.
The HBT radii, $R_{\mathrm side}$ and $R_{\mathrm out}$ corresponding to $q_{\mathrm side}$ 
and $q_{\mathrm out}$ extracted from the  (Gaussian) parametrization of $C_2$.
% is plotted in Fig.~\ref{fig1} (left panel).

%%%%%%%%%%%%%%%%%%%%%%%%%%    Fig. 2 %%%%%%%%%%%%%%%%%%%%%%%%%%%%%
\begin{figure}
\begin{center}
\includegraphics[scale=0.25]{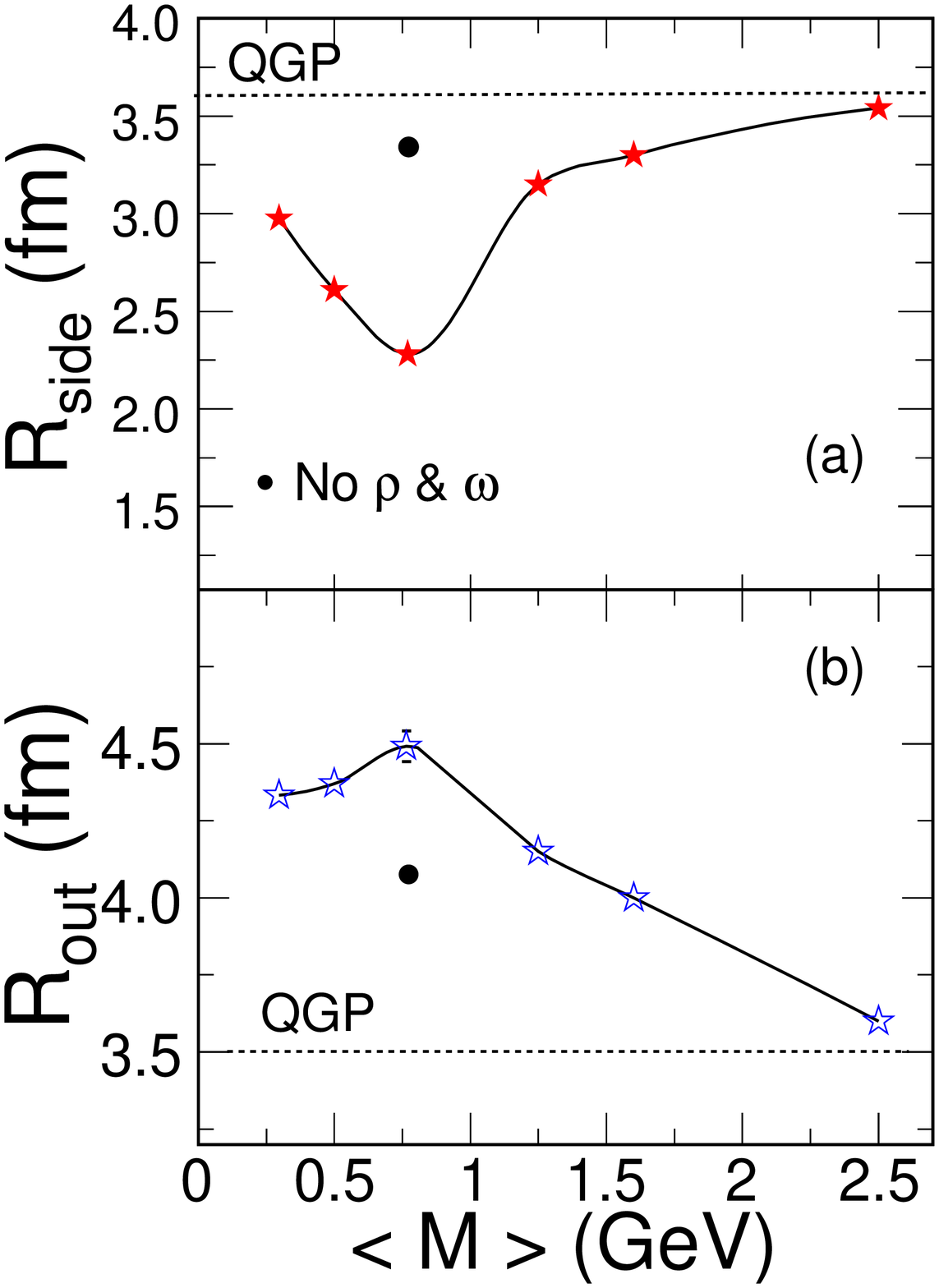}
\includegraphics[scale=0.4]{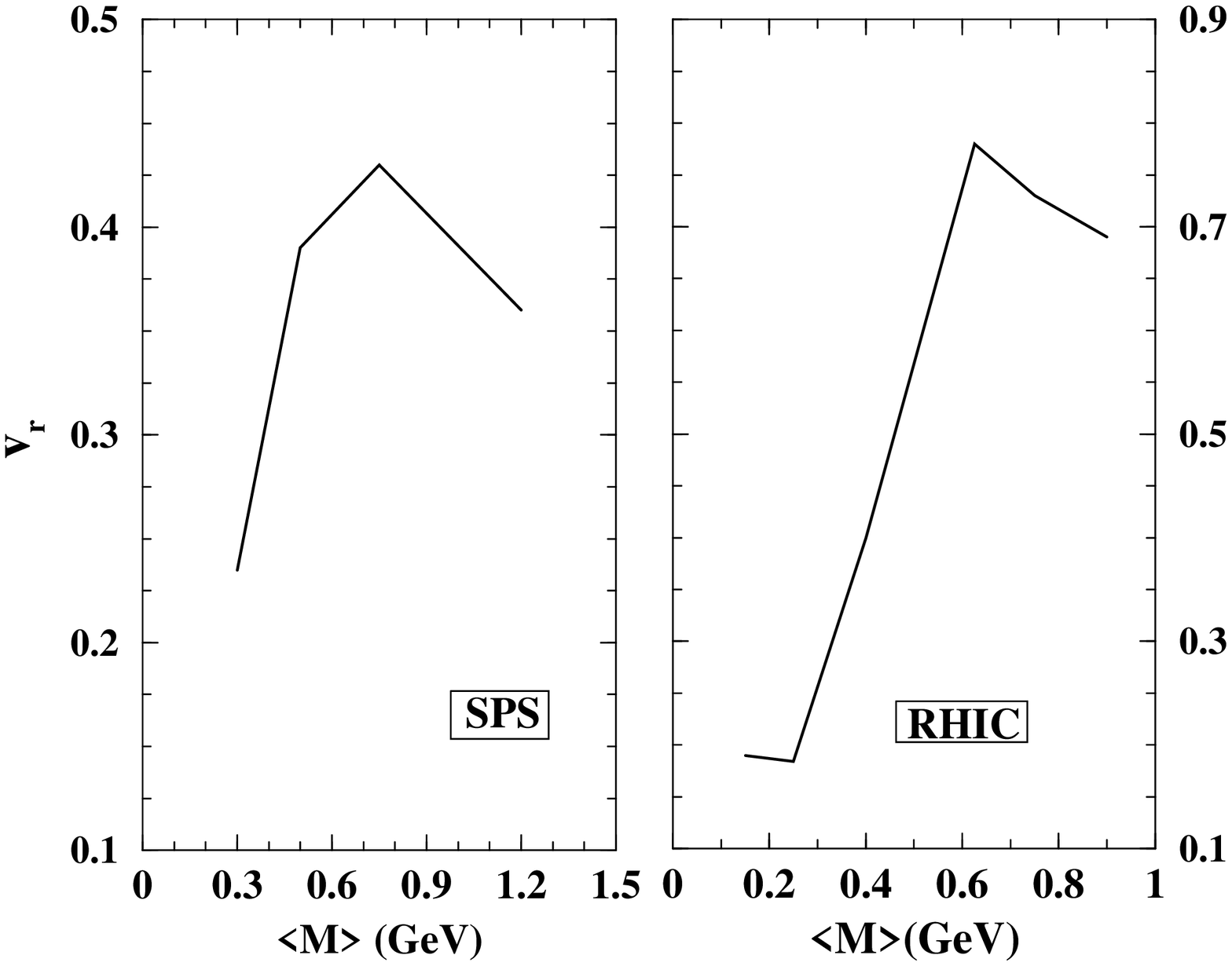}
\caption{Left panel: variation of 
$R_{side}$ and $R_{out}$ as a function of $\langle M \rangle$. 
The dashed (solidi with asterisk) line indicates HBT radii for 
QGP (total=QGP+hadron) phase.
Right panel: variation of the radial velocity with  $\langle M \rangle$
extracted from the ratio of 
the $p_T$ distribution of photons to lepton pairs
(see~\cite{pm} for details).
}
\label{fig1}
\end{center}
\end{figure}
%%%%%%%%%% End of Fig.2 %%%%%%%%%%%%%%%%%%%%%%%%%%%%%

The $R_{\mathrm side}$ 
is related to the transverse size of the system 
%and considerably affected by the collectivity and the
whereas the $R_{\mathrm out}$ measures both the 
transverse size and duration of particle emission (~\cite{hbt} for review). 
The $R_{\mathrm side}$ shows non-monotonic dependence on $\langle M\rangle$
(Fig.~\ref{fig1}, left panel). It can be shown
that $R_{side}\sim 1/(1+E_{\mathrm collective}/E_{\mathrm thermal})$. 
The high $\langle M\rangle$ regions 
are dominated by the early partonic phase~\cite{hbt} 
where the collective flow has not been developed fully 
consequently the ratio of collective ($E_{\mathrm collective}$)
to thermal ($E_{\mathrm thermal}$) energies is small-
hence a larger $R_{\mathrm side}$ is obtained for large $M$.
In contrast, the lepton pairs with $M\sim m_\rho$ 
are emitted from the late hadronic phase where the 
collective flow or 
$1+E_{\mathrm collective}/E_{\mathrm thermal})$ is large, 
which is reflected as a dip
in $R_{\mathrm side}$ for $\langle M\rangle\sim m_\rho$. 
Thus the variation of $R_{\mathrm side}$
with $M$ can be used as an efficient tool to measure the
collectivity in various phases of matter. 
%The dip in $R_{\mathrm side}$ at $\langle M\rangle\sim
%m_\rho$ is due to the contribution dominantly from the hadronic phase.
%The dip, in fact vanishes if the contributions from $\rho$ and $\omega$ 
%is switched off~\cite{mam}. 
%(solid circle in Fig.~\ref{fig1}). 
We observe that by keeping the $\rho$ and $\omega$ contributions
and setting radial velocity, $v_r=0$, the dip in $R_{\mathrm side}$
vanishes, confirming
the fact that the dip is caused by the 
large radial flow of the hadronic matter.   
The $R_{\mathrm out}$ probes both the transverse dimension as well as the 
duration of emission and unlike $R_{\mathrm side}$, $R_{\mathrm out}$ does not
remain constant even in the absence of radial flow. The large 
$\langle M\rangle$ regions are
populated by lepton pairs from early partonic phase where the
effect of flow is small and the duration of emission is also
small - resulting in smaller values of $R_{\mathrm out}$. 
For lepton pair from $\langle M\rangle\sim m_\rho$ region  the flow is large
which could have resulted in a dip as in $R_{\mathrm side}$ in
this $M$ region. However, $R_{\mathrm out}$ probes the duration
of emission too, which is large for hadronic phase.
The larger duration overwhelms
the reduction of $R_{\mathrm out}$
due to flow in the hadronic phase
resulting in a bump in $R_{\mathrm out}$ in this region of $\langle M\rangle$.

As mentioned before the $v_r$ can be
estimated from the ratio of the $p_T$ spectra of real photons to lepton
pairs. 
Fig.~\ref{fig1} (right panel) shows the variation of $v_r$
with $\langle M\rangle$ both for SPS and RHIC conditions. 
The individual spectra of photons and lepton pairs 
are constrained by the available experimental data~\cite{pm}.
The $v_r$ 
increases with M up to $M=M_{\rho}$ then drops.
%How can we understand this behaviour?
From the invariant mass spectra it is known that the low
$M$ (below $\rho$ mass)  and
high M (above $\phi$ peak) pairs originate from a partonic source~\cite{pm}.
The collectivity (or flow) does not develop fully 
in the QGP  resulting in smaller values of $v_r$ at both low 
and high $M$ regions.
Lepton pairs
for $M\sim m_\rho$ originate from the late hadronic source
which are largely affected
by the flow resulting in higher values of $v_r$.
In summary, the value of $v_r$ for $M$ below and above the
$\rho$-peak is small but around the $\rho$ peak is large
- the resulting behaviour is displayed in Fig.~\ref{fig1} (right panel).
Similar non-monotonic variation of the effective slope parameter 
of the $p_T$ distribution of lepton pairs for various $M$ windows
is observed in~\cite{sabya}. The evolution of $v_r$ as observed in
Fig.~\ref{fig1} (right panel) is responsible for such behaviour.

\section{Elliptic flow probed by single electron spectra}
The heavy flavors, namely, charm and bottom quarks, play a 
crucial role in characterizing the QGP (see also~\cite{rappicpa}).
As the relaxation time is larger for heavy quarks 
than light partons, the light quarks and the gluons thermalize 
faster.  Therefore, the  propagation of  heavy quarks through 
QGP may be treated as the 
interactions between equilibrium and non-equilibrium degrees
of freedom and the FPE provides an appropriate
framework~\cite{theory} for such studies. In this work we would like
to evaluate $v_2^{HF}$  and $R_{\mathrm AA}$ of 
heavy flavours within the framework of FPE and 
contrast the results with the available experimental data.
The evolution of heavy quarks momentum distribution function, while propagating
through the QGP are assumed to be governed by the FPE,  which reads,

\begin{eqnarray}
\frac{\partial f}{\partial t} = 
\frac{\partial}{\partial p_i} \left[ A_i(p)f + 
\frac{\partial}{\partial p_j} \lbrack B_{ij}(p) f \rbrack\right] 
\label{expeq}
\end{eqnarray}
where the kernels $A_i$ and $B_{ij}$ are given by,
$A_i =\int d^3 k\omega(p,k) k_i \,\,\,\,\, B_{ij}=\int d^3 k\omega(p,k) k_ik_j$,
for $\mid\bf{p}\mid\rightarrow 0$,  $A_i\rightarrow \gamma p_i$
and $B_{ij}\rightarrow D\delta_{ij}$, where $\gamma$ and $D$ stand for
drag and diffusion co-efficients respectively.

\begin{figure}
\begin{center}
\includegraphics[scale=0.3]{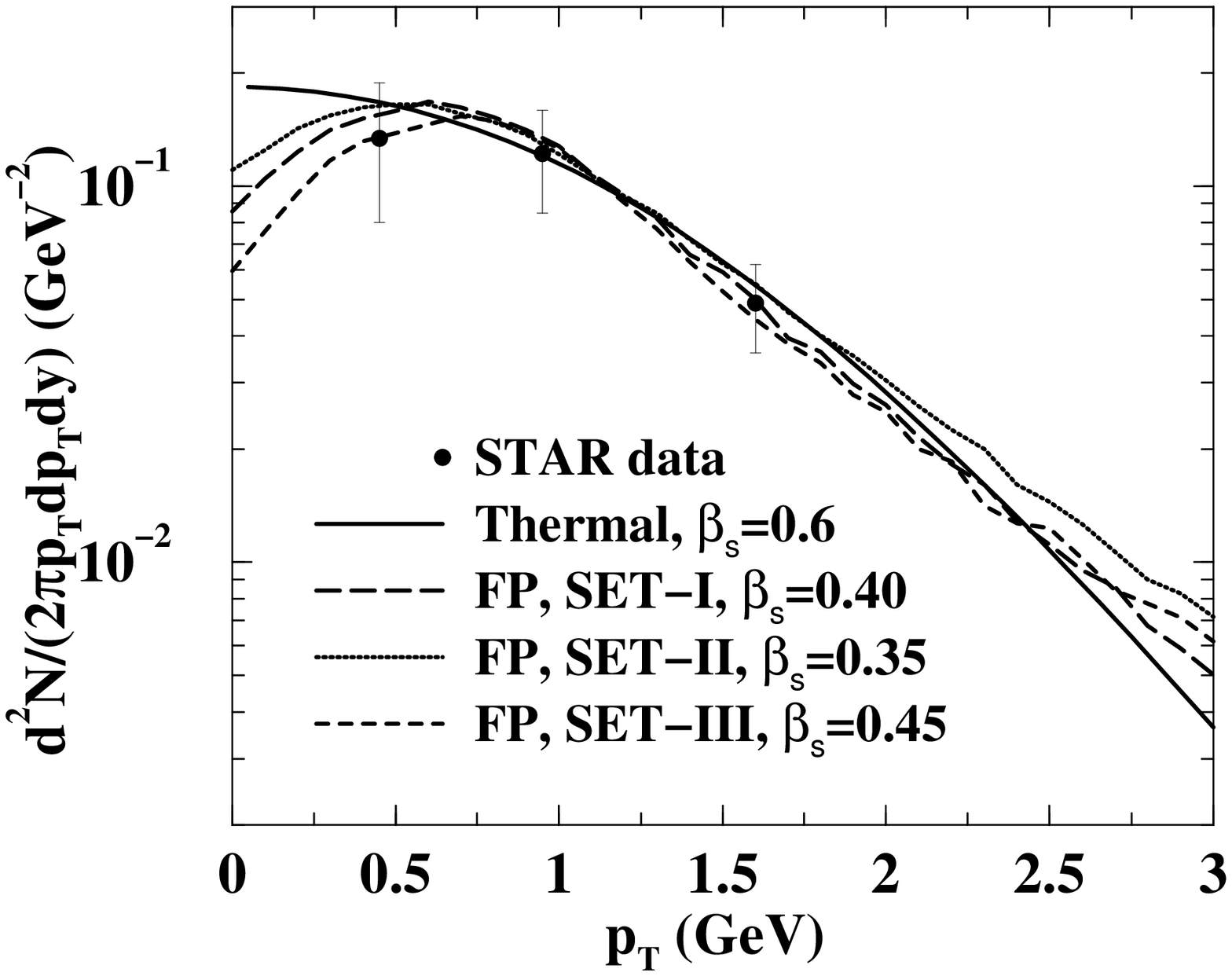}
\includegraphics[scale=0.3]{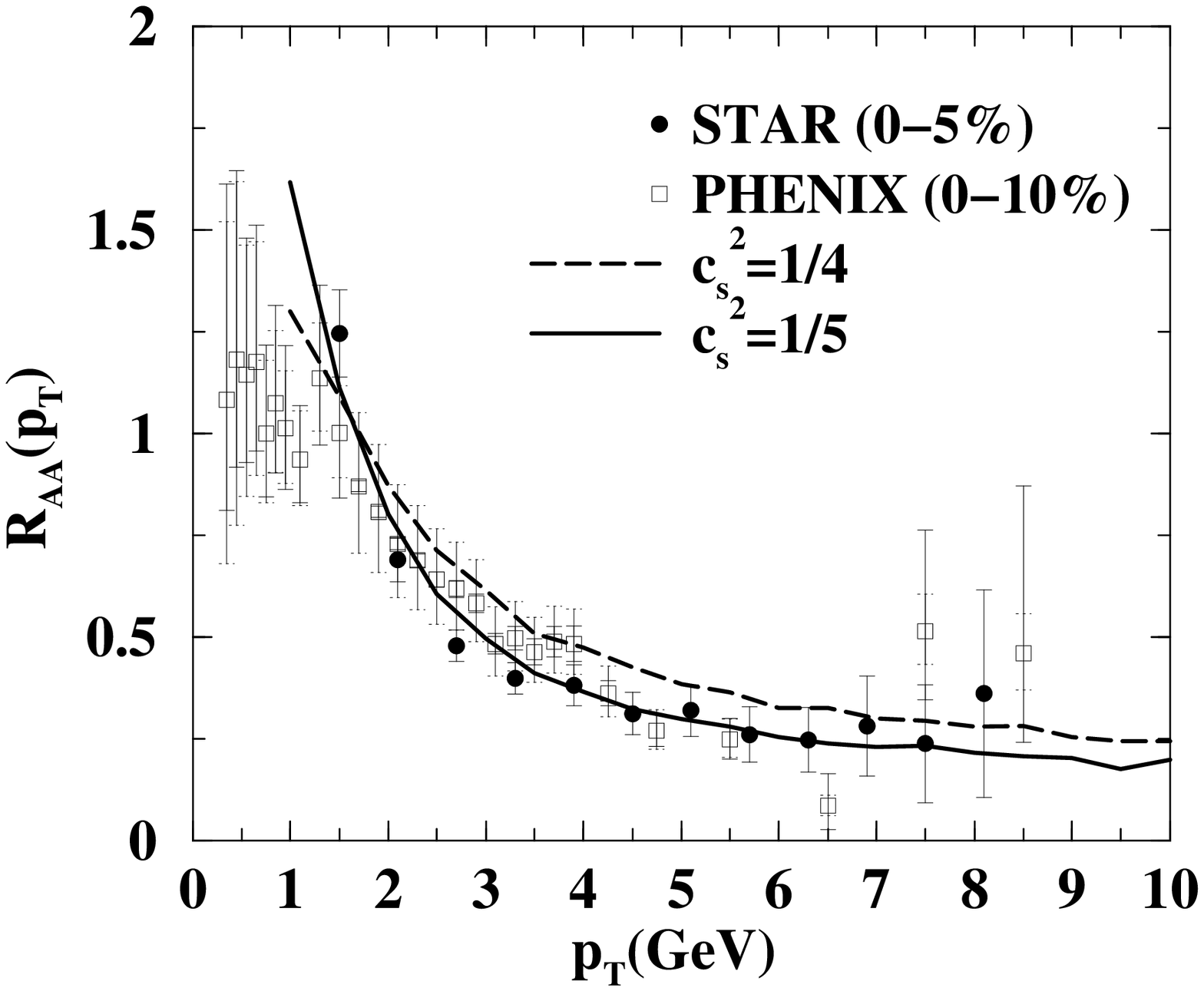}
\caption{Left panel: $p_T$ distribution of $D$ mesons. 
$\beta_s$ indicates the value of $v_r$ which appears as a
parameter in the blast wave model. FP 
stands for the results obtained from the solution of FP equation.
The experimental data from STAR collaboration~\cite{stard} is compared
with the theoretical results.  Right panel:
variation of $R_{\mathrm AA}$ with $p_T$. 
The initial temperature and thermalization time are
taken as 400 MeV and 0.2 fm/c respectively.
}
\label{fig2}
\end{center}
\end{figure}

The basic inputs required for solving the FP equation
are the dissipative co-efficients and initial momentum 
distributions of the heavy quarks. The (effective) drag and diffusion
coefficients have been evaluated by taking in to account
both the collisional and radiative processes~\cite{das2}.
In the radiative process the dead cone and LPM effects are included.
In evaluating the drag co-efficient we have
used temperature dependent  strong coupling,
$\alpha_s(T)$~\cite{Kaczmarek}.
The Debye mass, $\sim g(T)T$ also a temperature dependent
quantity used as  a cut-off to shield the infrared divergences
arising due to the exchange of massless gluons.
The initial momentum distribution of heavy quarks 
in pp collisions have been taken from the
NLO MNR~\cite{MNR} code. The solution of the FPE for the heavy  
quarks is convoluted with the fragmentation functions of the 
heavy quarks to obtain the $p_T$ distribution of the 
$D$ and $B$ mesons. For heavy-quark fragmentation function, the
Peterson function has been used. 
The solution of the FPE
has been used to predict the $p_T$ spectra of the $D$ mesons by 
following the procedure similar to blast wave method~\cite{das1},
the result is compared with experimental data
~\cite{stard} (Fig.~\ref{fig2}:left panel) 
which indicate that
the present data can not distinguish between the equilibrium and 
non-equilibrium scenario.  
The $p_T$ distribution 
of the electrons from the semi-leptonic decays of $D$ and $B$ meson
are evaluated using the standard techniques available in 
the literature. The ratio of the $p_T$ distribution of the 
electron from the decays of heavy flavours  produced  in heavy ion collisions 
to the corresponding (appropriately scaled by the number of collisions) 
quantities from the pp collisions is defined as:
$R_{\mathrm AA}(p_T)=dN^{e\,\,Au+Au}/d^2p_Tdy/
\left[N_{\mathrm coll}\times dN^{e\,\,pp}/d^2p_Tdy\right]$,
which will be unity in the absence of re-scattering.
The STAR~\cite{stare} and the PHENIX~\cite{phenixe} collaborations have
measured the $R_{\mathrm AA} (p_T)$ 
for non-photonic single electron as a function of
$p_T$ for Au+Au at $\sqrt{s_{NN}}=200$ GeV. 
The experimental data from both the collaborations
show $R_{AA}<1$ for $p_T\geq 2$ GeV indicating
substantial  interaction of the heavy quarks with
the plasma particles. The spectra evaluated using the formalism
described above reproduces the data reasonably well (Fig.~\ref{fig2}, right
panel).
%It is clear from the results displayed in Fig.~\ref{fig1} that
%the charm quarks suffer more suppression than bottom quarks at
%RHIC energy.

%%%%%%%%%%%%%%%%%%%%%%%%%%%%%%%%%%%%
%%%%%%%%%%%%%%%%%%%%%%%%%%%%%%%%%%%%%%%

\begin{figure}
\begin{center}
\includegraphics[scale=0.3]{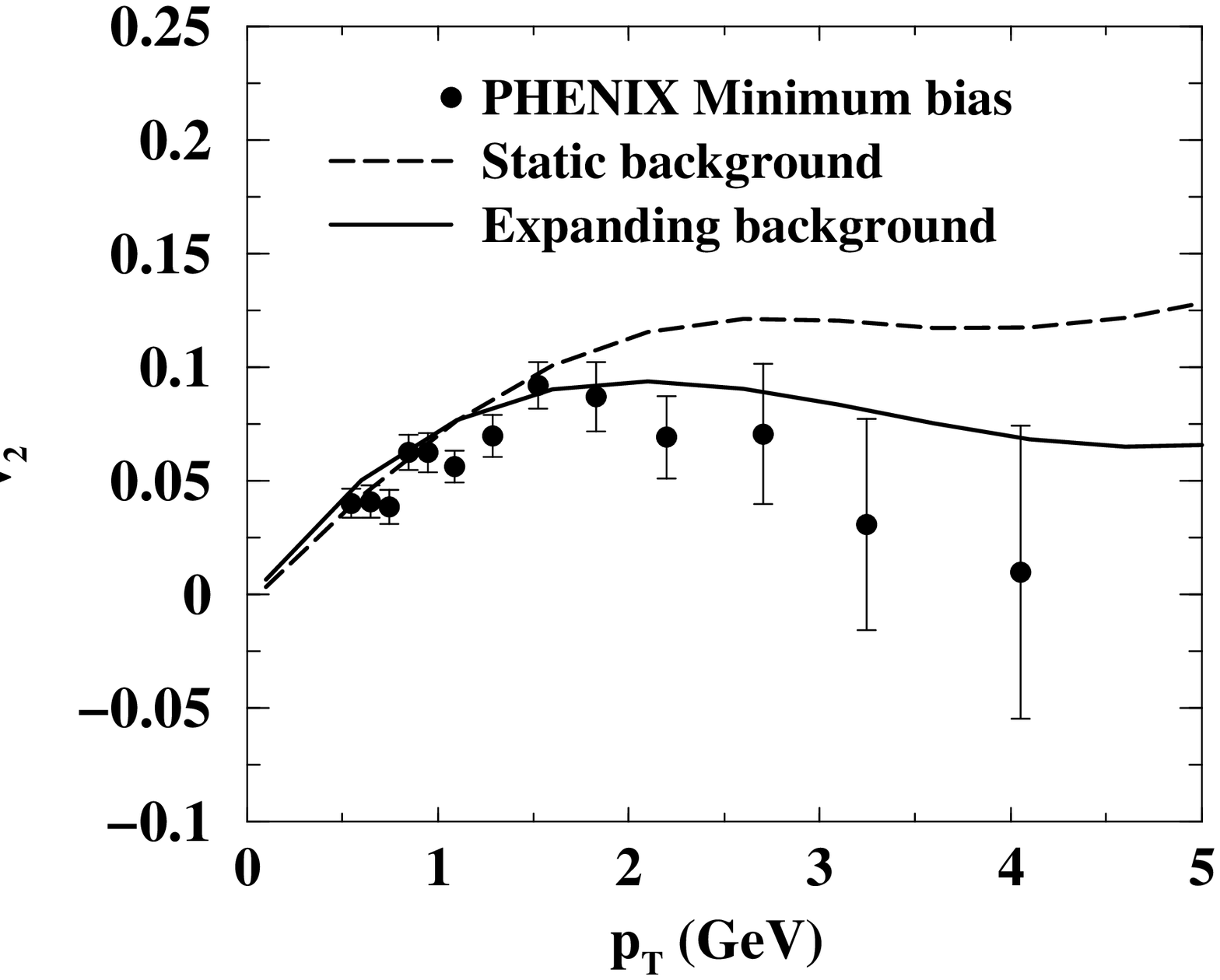}
\includegraphics[scale=0.35]{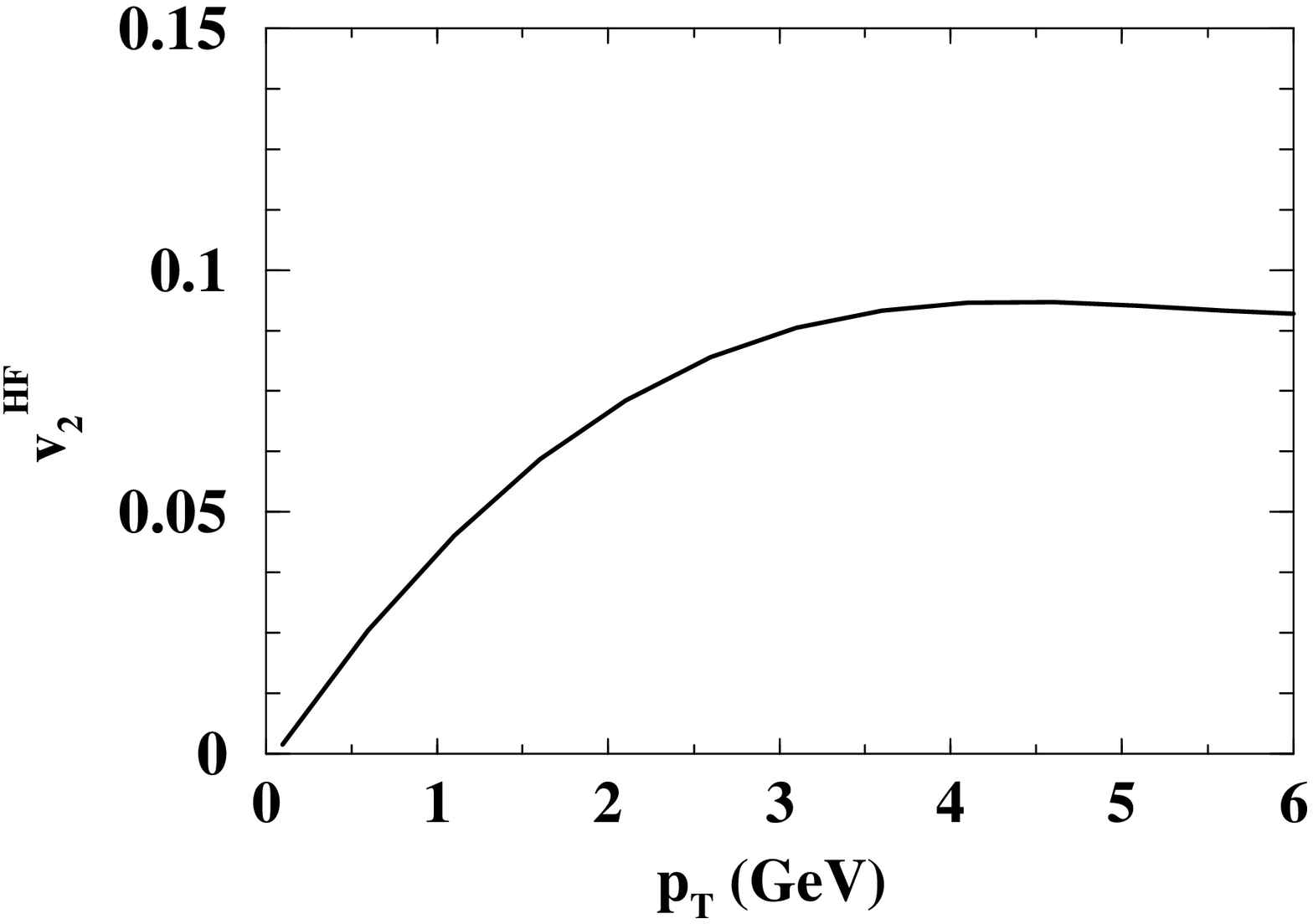}
\caption{Left panel: variation of $v_2^{HF}$ with $p_T$ for 
for RHIC energy. Right panel: $v_2^{HF}$ is plotted as a function of
$p_T$ for LHC energy for 0-10\% centrality. The values of $T_i$ and
$\tau_i$ are taken as  700 MeV and 0.08 fm/c respectively.
}
\label{fig3}
\end{center}
\end{figure}

\begin{figure}
\begin{center}
\includegraphics[scale=0.3]{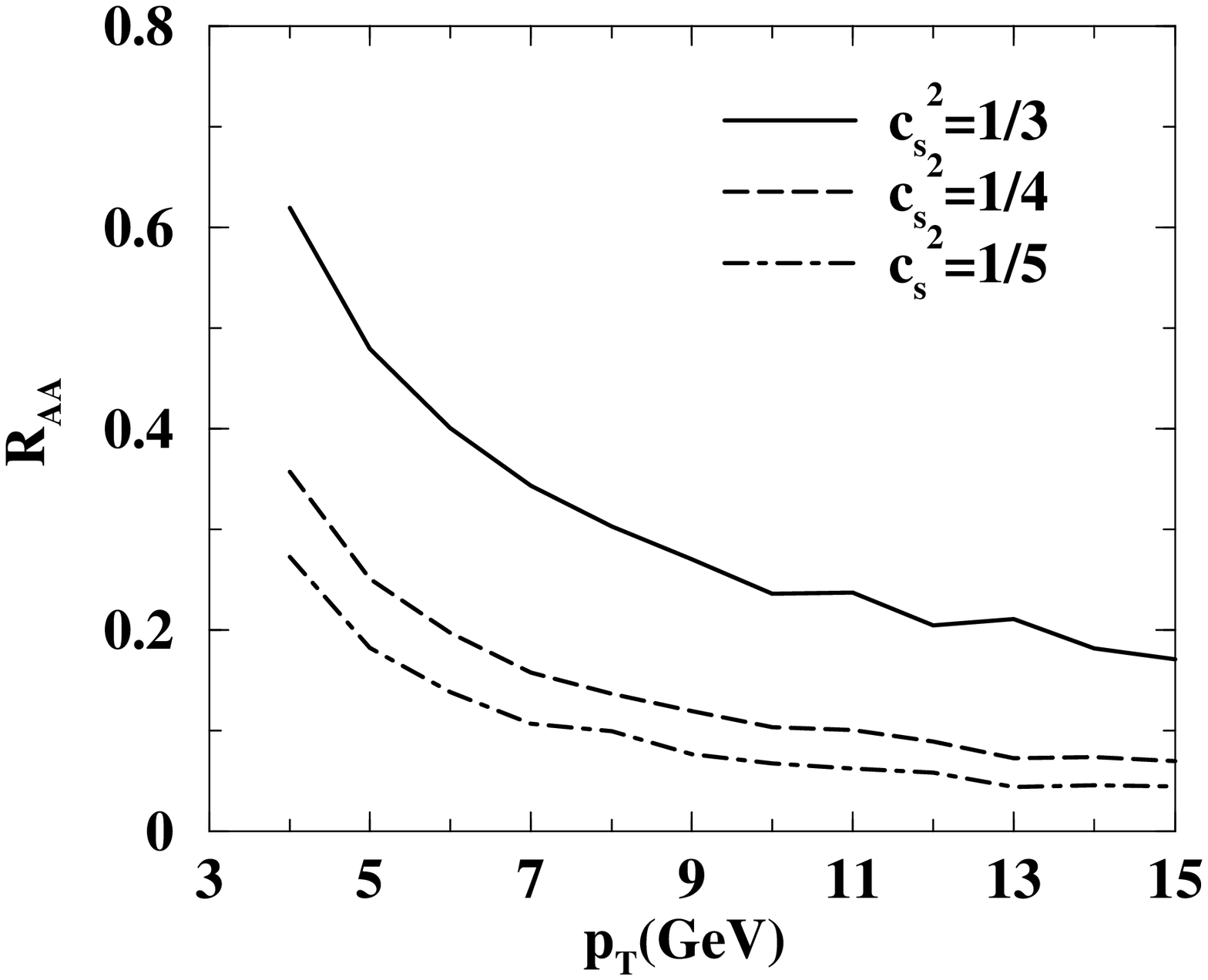}
\includegraphics[scale=0.3]{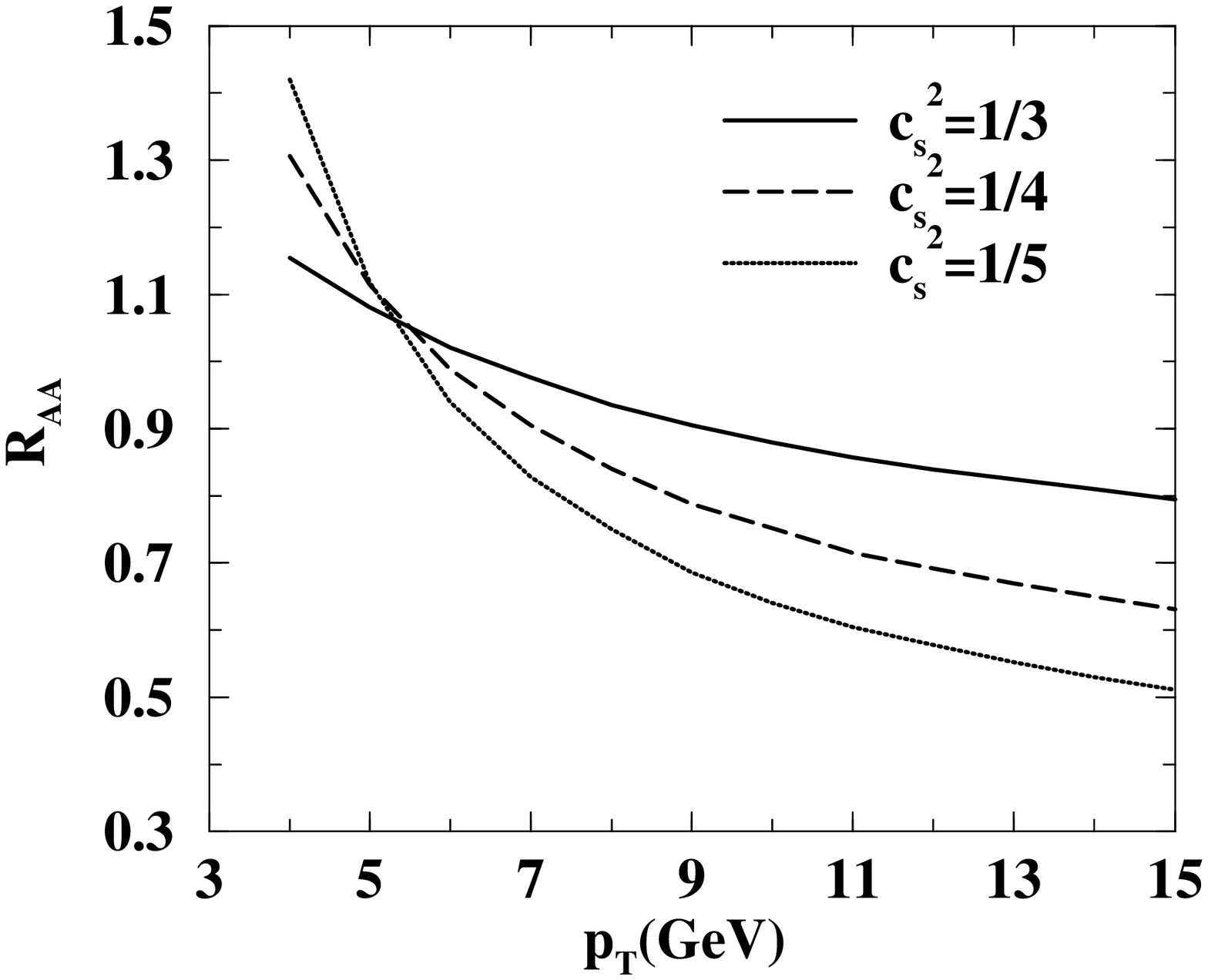}
\caption{Variation of $R_{\mathrm AA}$ 
with $p_T$ for LHC for 0-10\% centrality. Left panel
for charm and right panel bottom quarks. The values of $T_i$ and $\tau_i$
are taken as 700 MeV and 0.1 fm/c respectively. 
}
\label{fig4}
\end{center}
\end{figure}

Next we discuss the elliptic flow  resulting from non-central
collisions of nuclei.
When a heavy quark propagates along the major
axis of an ellipsoidal domain of QGP (resulting from the non-central collisions)
then the number of interactions it encounters or in other words
the amount of energy it dissipates or the amount of momentum degradation
that takes place is different from when it propagates
along the minor axis. 
Therefore, the momentum distribution of electrons
originating from the decays of heavy flavoured hadrons produced
from the fragmentation of heavy quarks  propagating through
an anisotropic domain will reflect such anisotropy. 
%Thus the spatial anisotropy due to non-central collisions
%will be reflected through  the momentum distributions.  
The degree
of momentum anisotropy will depend on both the spatial anisotropy and
more importantly on the coupling strength of the interactions 
between the heavy quarks and the QGP.  
The drag and diffusion coefficients depend on the temperature of
the background medium (QGP) which evolves in space and time due 
to expansion. Therefore, the drag and diffusion  will also change
due to the flow of the background.
The flow of the background has been treated
within the ambit of (2+1) dimensional hydrodynamics~\cite{hydro}.
%The FP equation describes the evolution of the heavy quarks
%due to its interaction with the QGP which undergoes hydrodynamic 
%flow~\cite{hydro}.
The coefficient of elliptic 
flow, $v_2^{HF}$ is defined as:
$v_2^{HF}(p_T)=\langle cos(2\phi) \rangle= 
\int d\phi dN/dydp_Td\phi|_{y=0} cos(2\phi)/\left[\int d\phi  
dN/dydp_Td\phi|_{y=0}\right]$.
We evaluate $v_2^{HF}$ in the current formalism~\cite{dasv2} 
and compare the results with experimental data~\cite{phenixemb}
(Fig.~\ref{fig3},  left panel). 
%%%%%%%%%%%%%%%%%%%%%%%%%%%%%%%%%%%%
%%%%%%%%%%%%%%%%%%%%%%%%%%%%%%%%%%%%%%%
The prediction for the elliptic
flow of the heavy quarks to be measured at Large Hadron Collider (LHC) energy 
through the semi-leptonic decays is depicted in Fig.~\ref{fig3} (right panel). 
The value of $v_2^{HF}$ at LHC is similar to that of at RHIC. 
The prediction for the $R_{\mathrm AAA}$ at LHC has been displayed in 
Fig.~\ref{fig4} separately for $D$ and $B$ mesons. 
The sensitivity of the results on the equation of state (EoS),
i.e. on the velocity of sound is also considered.
Lowering of $c_s$ gives more suppressions as observed in Fig.~\ref{fig4}. 
Lower value of velocity of sound,  $c_s$ makes the expansion of the plasma 
slower enabling the propagating heavy quarks to spend more time to interact
in the medium and hence lose more energy before exiting from the plasma
resulting in more suppression.

%%%%%%%%%%%%%%%%%%%%%%%%%%%%%%%%%%%%
%%%%%%%%%%%%%%%%%%%%%%%%%%%%%%%%%%%%%%%
\section{Summary} 
We have shown that the variation of various HBT radii with invariant mass
extracted from the correlation functions of two lepton pairs
can be used to understand the evolution of collectivity in 
ultra-relativistic heavy ion collision.
The evolution of the radial flow in the system produced in 
URHIC has been discussed and demonstrated that the non-monotonic variation 
of $v_r$ with $M$ can be used as a signal for parton to hadron transition. 
The elliptic flow of the matter probed by the single electrons originating
from the heavy flavour decays has been studied. The elliptic flow 
and the nuclear suppression factor measured at RHIC have been reproduced
and predictions for LHC have been given 
including both the radiative and the collisional processes of energy loss 
in evaluating the effective drag and diffusion coefficients.  
The  sensitivity of the results on the EoS has also been studied. 

%%%%%%%%%%%%%%%%%%%%%
{\bf Acknowledgment:} The author is grateful to Santosh K Das, Sabyasachi Ghosh, Bedangadas Mohanty,
Payal Mohanty, Jajati K Nayak and Sourav Sarkar for collaboration and 
to Tetsufumi Hirano for providing hadronic chemical potentials. 
This work is partially supported by DAE-BRNS Sanction No.  2005/21/5-BRNS/2455.

\end{document}